\documentclass[twocolumn,showpacs]{revtex4}

\usepackage{graphicx}
\usepackage{dcolumn}
\usepackage{bm}
\usepackage{latexsym}

\begin{document}

\title{Are there ghosts in the self-accelerating brane universe?}

\author{Kazuya Koyama$^{1,2}$}
 
\affiliation{$^1$
Department of Physics, University of Tokyo 
7-3-1 Hongo, Bunkyo, Tokyo 113-0033, Japan\\
$^2$ Institute of Cosmology and Gravitation, Portsmouth 
University, Portsmouth, PO1 2EG, UK}

\date{\today}

\begin{abstract}
We study the spectrum of gravitational perturbations about a vacuum de Sitter 
brane with the induced 4D Einstein-Hilbert term, in a 5D Minkowski 
spacetime (DGP model). We consider solutions that include a self-accelerating
univese, where the accelerating expansion of the universe is 
realized without introducing a cosmological constant on the brane.
The mass of the discrete mode for the spin-2 graviton 
is calculated for various $Hr_c$, where $H$ is the Hubble parameter and $r_c$ is the cross-over 
scale determined by the ratio between the 5D Newton constant and the 4D Newton 
constant. We show that, if we introduce a positive cosmological constant on the brane
($Hr_c >1$), the spin-2 graviton has mass in the range $0 < m^2 < 2H^2$ and 
there is a normalisable brane fluctuation mode with mass $m^2=2 H^2$. 
Although the brane fluctuation mode is healthy, the spin-2 graviton has 
a helicity-0 excitation that is a ghost. If we allow a negative 
cosmological constant on the brane, the brane bending mode becomes a ghost for 
$1/2 < Hr_c <1$. This confirms the results obtained by the boundary effective action 
that there exists a scalar ghost mode for $Hr_c >1/2$.  
In a self-accelerating universe $Hr_c=1$, the spin-2 graviton has mass $m^2=2H^2$, 
which is known to be a special case for massive gravitons in de Sitter spacetime where 
the graviton has no helicity-0 excitation and so no ghost. 
However, in DGP model, there exists a brane fluctuation mode with the same mass 
and there arises a mixing between the brane fluctuation mode and the spin-2 graviton.
We argue that this mixing presumably gives a ghost in the self-accelerating 
universe by continuity across $Hr_c=1$, although a careful calculation of the effective action 
is required to verify this rigorously.
\end{abstract}

\maketitle
\section{Introduction}
The cosmological constant problem is one of the most difficult and 
important problems in particle physics and cosmology \cite{W}. 
The old problem is the smallness of the cosmological constant 
compared with the fundamental scales of particle physics. 
A natural solution for this problem has been thought to be 
an exact cancellation of the cosmological constant. However,   
recent discoveries of the accelerated expansion of the universe make the 
problem more complicated \cite{SN}. 
Within the 4D Einstein theory of gravity, we do need a tiny cosmological 
constant in order to explain the acceleration 
of the universe. Then we must introduce a tiny cosmological constant
by hand and accept a fine-tuning \cite{AD}. 

There have been many attempts to modify the 4D Einstein theory of gravity 
to explain the acceleration of the universe instead of introducing 
the cosmological constant \cite{T}.
One such attempt was made in the context of the brane world models, 
where our universe is realized as a 4D hypersurface (brane) in a 
higher dimensional spacetime (bulk). Dvali, Gabadadze and Porrati (DGP) 
proposed a model where the 4D Einstein-Hilbert term is 
assumed to be induced on the brane \cite{DGP}. In this model, the accelerated 
expansion of the universe can be realized by a 
modification of Einstein gravity on large scales and we do not need a 
cosmological constant (self-accelerating universe) \cite{D}. 

Recently several authors claimed that there exist ghost-like excitations
in the self-accelerating universe \cite{ghost,ghost1}. 
If this is true, it becomes difficult to consider the DGP model 
as a consistent model to explain the acceleration of the 
universe. In this paper, we re-examine the existence of 
the ghost in the DGP model based on a direct claculation 
of the spectrum of the perturbations.

\section{Background spacetime and pertrubations}

The 5D action describing the DGP model is given by
\begin{eqnarray}
S &=& \frac{1}{2 \kappa^2}\int d^5 x \sqrt{-g} R 
+ \frac{1}{2 \kappa_4^2} \int d^4 x \sqrt{-\gamma} \:\: 
{}^{(4)\!}R \nonumber\\
&& - \sigma \int d^4 x \sqrt{-\gamma} + \frac{1}{\kappa^2} 
\int d^4 x \sqrt{-\gamma}  K,
\end{eqnarray}
where $\sigma$ is the tension of the brane, $K_{\mu \nu}$ is the extrinsic curvature 
and $K=K^{\mu}_{\:\: \mu}$. In this letter, we only consider a vacuum energy 
contribution from matter fields on the brane for simplicity. 
We also assume reflection symmetry across the 
brane. Then the junction condition must be imposed at the 
brane as 
\begin{equation}
K^{\mu}_{\:\: \nu} = {\kappa^2 \over 2} 
\left( -{\sigma \over 3} \delta^{\mu}_{\:\: \nu}
+ {1 \over \kappa_4^{2}} 
{}^{(4)\!}\tilde{G}^{\mu}_{\:\: \nu} \right),
\end{equation}
where $^{(4)\!} G^{\mu}_{\:\: \nu}$ is the Einstein tensor on the brane and 
${}^{(4)\!}\tilde{G}^{\mu}_{\:\: \nu} ={}^{(4)\!}
G^{\mu}_{\:\: \nu}-(1/3)  {}^{(4)\!} G \delta^{\mu}_{\:\: \nu}$. 
The 4D Einstein tensor comes from the induced Einstein-Hilbert term.
The Friedmann equation on the brane is given by
\begin{equation}
\pm H = r_c H^2- \frac{\kappa^2}{6} \sigma, \quad r_c = \frac{\kappa^2}{2 \kappa_4^2}.
\end{equation}
The 5D solution for the metric with the 4D de Sitter brane can be obtained as \cite{D} 
\begin{equation}
ds^2 = dy^2 + N(y)^2 \gamma_{\mu \nu} dx^{\mu} dx^{\nu}, \quad
N(y)=1 \pm H y,
\end{equation}
where $\gamma_{\mu \nu}$ is the metric for the de Sitter spacetime
and the brane is located at $y=0$.
If we take the $+$ branch solutions, there is a solution for the de Sitter 
spacetime without $\sigma$,
\begin{equation}
H= \frac{1}{r_c}.
\end{equation}
We call this solution the self-accelerating universe.

Let us investigate the perturbations $N(y)^2 \gamma_{\mu \nu} + h_{\mu \nu}$ 
about the background de Sitter spacetime. 
In the following, we assume $H r_c \neq 1$ and treat the case $Hr_c=1$ 
separately. In addition to the 
gravitational perturbations $h_{\mu \nu}$, we must take into account a
perturbation of the position of the brane $y=\varphi(x)$ \cite{Tanaka}. 
Using the transverse-traceless gauge $\nabla^{\mu} h_{\mu \nu}=h=0$, 
the perturbed junction condition is given by
\begin{eqnarray}
k_{\mu \nu} &-& {\cal H} h_{\mu \nu} 
- r_c \left[ X_{\mu \nu}(h) 
-\kappa_4^2 \left(T_{\mu \nu} - \frac{1}{3} \gamma_{\mu \nu} T \right)
\right]\nonumber\\
&=& -(1-2 {\cal H} r_c) 
\left(   
\nabla_{\mu} \nabla_{\nu} + H^2  \gamma_{\mu \nu} \right) \varphi,
\label{junction}
\end{eqnarray}
where $k_{\mu \nu} =(1/2) \partial_y h_{\mu \nu}$, 
${\cal H} = \partial_y N/N$ on the brane and
$X_{\mu \nu}$ is given by
\begin{eqnarray}
X_{\mu \nu} &=& \delta {}^{(4)} G_{\mu \nu} + 3 H^2 h_{\mu \nu} \nonumber\\
&=& -\frac{1}{2}\left(\Box_4 h_{\mu \nu} 
-\nabla_{\mu} \nabla_{\alpha} h^{\alpha}_{\nu} 
-\nabla_{\nu} \nabla_{\alpha} h^{\alpha}_{\mu} 
+ \nabla_{\mu} \nabla_{\nu} h \right) \nonumber\\
&-& \frac{1}{2} \gamma_{\mu \nu}(\nabla_{\alpha} \nabla_{\beta}
h^{\alpha \beta} - \Box_4 h)
+H^2 \left( h_{\mu \nu} +  \frac{1}{2} \gamma_{\mu \nu} h \right). \nonumber\\
\end{eqnarray}
The equation of motion for $\varphi$ is obtained from the traceless 
condition $h=0$;
\begin{equation}
(1 -2 {\cal H} r_c)(\Box_4 +4 H^2) \varphi \nonumber\\
= \frac{\kappa^2 T}{6}. 
\end{equation}

Let us find solutions for the vacuum brane $T_{\mu \nu}=0$. 
Using the separation of variables $h_{\mu \nu} = \int dm~ e_{\mu \nu}(x) F_m(y)$, 
the equation of motion in the bulk is written as 
\begin{equation}
F_m'' + \frac{1}{N^2} (m^2 -2H^2) F_m=0,
\end{equation}
where prime denotes a derivative with respect to $y$.
There are two types of solutions. One type of solution is 
an inhomogeneous solution sourced by the scalar mode $\varphi$. 
We call this solution the spin-0 perturbation \cite{spin}. 
The other solution is a homogeneous solution with $\varphi=0$,
which is called the spin-2 perturbation. The spin-2 perturbations 
$\chi_{\mu \nu}$ satisfy the junction condition without $\varphi$
\begin{equation}
\chi_{\mu \nu}'-2 {\cal H} \chi_{\mu \nu} = -m^2 r_c \chi_{\mu \nu}.
\end{equation}
We find a tower of continuous Kaluza-Klein (KK) modes 
starting from $m^2 = (9/4) H^2$ as well as a normalizable discrete mode 
$m_d^2 =0$ in the $-$ branch and with 
\begin{equation}
\frac{m_d^2}{H^2} =\frac{1}{(Hr_c)^2}(3 Hr_c -1),
\end{equation} 
in the $+$ branch for $Hr_c >2/3$ \cite{KK}. 
For $H r_c >1$, the mass is in the range $0 < m_d^2 \leq 2H^2$ where $m_d^2 = 2H^2$ for 
the self-accelerating universe $Hr_c=1$ and $m_d^2 \to 0$ for $H r_c \to \infty$
\cite{comment}. 

In the $-$ branch, there are no 
normalizable solutions for the spin-0 perturbations. In the $+$ branch, there is
a normalizable solution given by
\begin{equation}
h_{\mu \nu} = \frac{1-2Hr_c}{H(1-Hr_c)} (\nabla_{\mu} \nabla_{\nu} 
+ H^2 \gamma_{\mu \nu}) \varphi.
\label{solphi}
\end{equation}
This is a solution with $m^2 = 2H^2$. 
~~
\section{Boundary effective action}

We can construct the 2nd order action for $h_{\mu \nu}$ and $\varphi$
from the 5D action by extending the result of Ref.\cite{GS}. The 
result is given by
\begin{equation}
\delta_2 S = -\frac{1}{4 \kappa^2} \int d^5 x \sqrt{-g} N^{-4} h^{\mu \nu} 
\delta {}^{(5)} G_{\mu \nu} 
+\frac{1}{\kappa^2} \int d^4 x\sqrt{-\gamma} {\cal L}_B,
\label{5Dactionp1}
\end{equation}
where $\delta {}^{(5)} G_{\mu \nu}$ is the 5D perturbed Einstein 
tensor and 
\begin{eqnarray}
{\cal L}_B &=& k^{\mu \nu} h_{\mu \nu} -k h
+ \frac{1}{2} {\cal H} (h^2 - h^{\mu \nu} h_{\mu \nu})\nonumber\\ 
&+& (1-2 {\cal H} r_c) 
\left(h_{\mu \nu} \nabla^{\mu} \nabla^{\nu} \varphi - h 
\nabla^{\rho} \nabla_{\rho} \varphi -3 H^2 h \varphi \right)
\nonumber\\
&-& 3{\cal H} \left(-(1-2 {\cal H} r_c)\varphi (\Box_4 + 4H^2) \varphi 
+ \frac{\kappa^2}{3} T \varphi \right) \nonumber\\
&+& \frac{1}{2} \kappa^2 h^{\mu \nu} T_{\mu \nu}  
-\frac{r_c}{2} h^{\mu \nu} X_{\mu \nu}(h).
\label{5Dactionp2}
\end{eqnarray}
This action gives the correct equation of motion and 
the junction condition for $h_{\mu \nu}$ and the equation of 
motion for $\varphi$. 

We can derive an effective action for the
brane fluctuation $\varphi$ by substituting the 5D solution for $h_{\mu \nu}$
given by $\varphi$ (\ref{solphi}) into the 5D action and get the off-shell 
action for $\varphi$ by integrating out only with respect to 
the extra coordinate $y$ \cite{Padilla}.
This yields the action for $\varphi$ in the $+$ branch as
\begin{equation}
S_{\varphi} = \frac{3 H}{2 \kappa^2} 
\left(\frac{1- 2Hr_c}{1- H r_c}\right) 
\int d^4 x \sqrt{-\gamma} \varphi (\Box_4 + 4 H^2) \varphi.
\label{actphi}
\end{equation}
We find that, for $Hr_c >1$, the kinetic term is always 
positive, so the brane fluctuation mode $\varphi$ is not a ghost.

However, there is a problem for the spin-2 perturbations. 
We consider only the + branch solutions that include the 
self-accelerating universe.
The 4D effective action for the spin-2 perturbations is also obtained
in a similar way. For the discrete mode with $m_d^2$, we get
\begin{equation}
S_{\chi} = \frac{r_c (3 Hr_c-1)}{4 \kappa^2 (3 Hr_c-2)} 
\int d^4x \sqrt{-\gamma} \chi^{\mu \nu} (\Box_4 -2H^2 - m_d^2) \chi_{\mu \nu},
\label{actchi}
\end{equation}
where transverse-traceless gauge fixing conditions 
$\nabla^{\mu} \chi_{\mu \nu} =\chi^{\mu}_{\nu}=0$ are imposed . 
This is exactly the same action for the 
spin-2 perturbations in the 4D massive gravity theory 
where the Pauli-Fierz (PF) mass term is added to the Einstein-Hilbert action 
by hand \cite{PF} 
\begin{equation}
S_M = -\frac{M^2}{8 \kappa_4^2}\int d^4 x \sqrt{-\gamma}(h^{\mu \nu} h_{\mu \nu} - h^2).
\end{equation}
Note that in the 4D PF theory, the mass term breaks the gauge symmetry for the 
perturbation and the transverse-traceless 
conditions are imposed by the equation of motion as constraints. 
In the limit $Hr_c \to \infty$, the action approaches 
the one for massless spin-2 perturbations. However, there is a 
discontinuity between the massless perturbations and the massive 
perturbations that is known as the van Dam-Veltman-Zakharov 
discontinuity \cite{vDVZ}. 
Due to the lack of gauge symmetry, the massive
spin-2 perturbations contain a helicity-0 excitation. Moreover, 
it has been shown that this helicity-0 excitation becomes a ghost
if $0 < M^2 < 2H^2$ \cite{Higuchi,DW}. 
This is exactly the same mass range for the 
discrete mode in the $+$ branch 
for $Hr_c > 1$. 
Thus we expect that there appears a ghost 
when $Hr_c > 1$, that is, the universe with a positive cosmological
constant is unstable. 

\section{4D effective action for helicity-0 excitations}

In summary,  
we expect that the helicity-0 excitation for the spin-2 perturbations is a ghost
for $Hr_c > 1$ in the $+$ branch.
In order to confirm this result, we construct 
the effective action for the helicity-0 excitations in the DGP model. 
It is convenient to take the metric as 
\begin{eqnarray} 
ds^2 &=& (1+2A_{yy})dy^2  
+2 N(y) A_y   dt dy  \nonumber\\
&+& N(y)^2 \left( -(1+2 A )dt^2  
+ a^2 (1+2{\cal R} )\delta_{ij} dx^i dx^j  \right), 
\nonumber
\end{eqnarray} 
where $a(t)=\exp(Ht)$. 
In the absence of bulk matter, perturbations are solved 
using a ``master variable'', $\Omega$ \cite{M}. 
In the special case of a de Sitter brane in a Minkowski bulk, the 
metric variables are written via the master variable $\Omega$ as 
%
\begin{eqnarray} 
{A} &=& -\frac{1}{6 a} \biggl( 2\Omega'' 
-\frac{N'}{N} \Omega' + \frac{1}{N^2} {\ddot\Omega} \biggr), 
\label{metric_omega_phi} 
\nonumber\\ 
{A_y} &=& \frac{1}{a N} \biggl( {\dot\Omega}' - \frac{N'}{N} {\dot\Omega} 
\biggr), 
\label{metric_omega_s} 
\nonumber\\ 
{A_{yy}} &=& \frac{1}{6 a} \biggl( \Omega'' 
-2 \frac{N'}{N} \Omega' +\frac{2}{N^2} {\ddot\Omega} \biggr), 
\label{metric_omega_n} 
\nonumber\\ 
{R} &=& \frac{1}{6 a} \biggl( \Omega'' +\frac{N'}{N} \Omega'
-\frac{1}{N^2} {\ddot\Omega} \biggr),
\label{metric_omega_psi} 
\end{eqnarray} 
where dot denotes a derivative with respect to $t$.
In the bulk, the master equation for $\Omega$ can be 
solved by the separation of variables 
$\Omega = \int dm~ g_{mk}(t) f_m(y) e^{i k x}$ and 
the equation of motion in the bulk is given by
\begin{eqnarray} 
f_m''-2 \frac{N'}{N} f_m' + \frac{m^2}{N^2} f_m &=& 0, \nonumber\\
\ddot{g}_{mk} -3H \ddot{g}_{mk} + \frac{k^2}{a^{2}} g_{mk} &=& -m^2 g_{mk}.
\label{eq_omega} 
\end{eqnarray} 
The junction conditions for $\Omega$ were derived in Ref.\cite{D1, D2}. 
For the vacuum brane, the boundary condition for $\Omega$ is given by
\begin{equation}
\Omega'-H \Omega - \frac{r_c}{1-2Hr_c} (2H^2 -m^2) \Omega=0.
\end{equation}
We again find a tower of continuous massive modes starting from 
$m^2=(9/4) H^2$, 
which are the KK modes for the spin-2 perturbations. In addition, 
there are two discrete modes. One is the mode with $m^2=2H^2$,
which is the spin-0 perturbation and the other is the mode 
with $m^2=m_d^2$, which is the helicity-0 excitation of 
the spin-2 perturbations. The 2nd order action for the 
master variable can be calculated by extending the result of 
Ref.\cite{YK}. Then we can construct the 4D effective action 
for the discrete modes with mass $m_i^2$ by integrating out with 
respect to $y$ as
\begin{equation} 
S_i= \frac{k^4 {\cal N}_{m^2_i}}{6 H \kappa^2} \int d^4 x a^{3} 
\psi_{mk} \left( \Box_4 -m_i^2 \right) \psi_{mk},
\label{action_Omega} 
\end{equation}  
where
\begin{equation}
{\cal N}_{m_i^2}= \frac{H r_c}{1-2Hr_c} 
+\frac{1}{2 \sqrt{\frac{9}{4}-\frac{m_i^2}{H^2}}},
\end{equation}
and $\psi_{mk}=a^{-3} g_{mk}$. 
We should note that there is no mixing between two discrete modes
for $H r_c \neq 1$. 
For the spin-0 perturbation 
with $m^2 =2 H^2$, ${\cal N}_{m_i^2}$ becomes
\begin{equation}
{\cal N}_{2H^2}=\frac{1-Hr_c}{1-2Hr_c}.
\label{radiona}
\end{equation}
${\cal N}_{2H^2}$ is positive for $H >1/r_c$, so
it is not a ghost. 
For the spin-2 perturbation with $m^2 = m_d^2$, 
${\cal N}_{m_d^2}$ becomes
\begin{equation}
{\cal N}_{m_d^2} = - \left(\frac{1-Hr_c}{1-2Hr_c} \right) 
\left( \frac{Hr_c}{3Hr_c-2} \right).
\label{spin2a}
\end{equation}
${\cal N}_{m_d^2}$ is negative for $H > 1/r_c$, which confirms
that the spin-2 perturbations contain a ghost.

\section{self-accelerating universe}
In a self-accelerating universe $Hr_c=1$, the spin-2 graviton
has a mass $m^2=2 H^2$. In the PF massive gravity theory 
with $M^2 = 2H^2$, the action is invariant under the transformation 
\cite{Higuchi, DW}
\begin{equation}
h_{\mu \nu} \to h_{\mu \nu} + (\nabla_{\mu} \nabla_{\nu} + H^2 \gamma_{\mu \nu})
\omega(x).
\label{gaugepf}
\end{equation}
Due to this symmetry, there are no physical helicity-0 excitations 
and no ghost. However,  
in the DGP model, the situation is different due to the 
existence of the brane fluctuation mode $\varphi$. 
For $Hr_c=1$, there is no longer a solution of the form Eq.~(\ref{solphi}) 
for the spin-0 mode due to the symmetry Eq.~(\ref{gaugepf}), but there is a
solution of the form $h_{\mu \nu} =A_{\mu \nu}(x) + B_{\mu \nu}(x)
\log N(y)$ where $A_{\mu \nu}$ and $B_{\mu \nu}$ are determined 
by $\varphi$ \cite{GSS}. Then there is a mixing between the spin-0 mode 
and the helicity-0 excitation of the spin-2 graviton, which can give a ghost 
\cite{GSS},\cite{R}. 
The derivation of the effective action is more subtle in this 
case and a careful calculation of the effective action is required. 

However, we can argue the existence of the ghost in the following way \cite{R}. 
From the effective action for the brane bending mode, it is clear that the brane 
bending mode becomes a ghost for $1/2 < H r_c <1$ Eq.~(\ref{radiona}). 
On the other hand, for $Hr_c >1$, the spin-2 graviton 
becomes a ghost (see FIG.~1). Thus, by continuity, it is likely to 
have a ghost for $H r_c=1$, in this case, from a mixing between the brane bending 
mode and the spin-2 graviton. It should be mentioned that this is consistent 
with the result obtained by the boundary effective action in Refs.\cite{ghost,ghost1}, 
where a scalar mode is found to be a ghost if $H r_c > 1/2$. 

\begin{figure}[t]
\centerline{
\includegraphics[width=9cm]{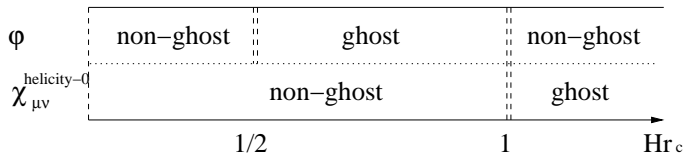}}
\caption{Summary of the existence of the scalar ghost for various $Hr_c$}
\label{fig:table}
\end{figure}

\section{Summary}
In this paper, we studied the spectrum of 
gravitational perturbations in the DGP model. 
In the self-accelerating branch, we showed that the spin-2 graviton
has a discrete mode with mass in the range $0 < m^2 \leq 2H^2$ for $Hr_c>1$ 
where $m^2 = 2H^2$ for the self-accelerating universe 
$Hr_c=1$ and $m^2 \to 0$ for $H r_c \to \infty$. 
Then the spin-2 graviton acquires a helicity-0 excitation
that is a ghost for $Hr_c>1$, 
which confirms earlier results \cite{ghost,ghost1}.
In addition, there is a normalisable brane fluctuation mode with mass 
$m^2=2 H^2$ that is not a ghost. 

The self-accelerating universe is special in the sense that the
spin-2 graviton has a mass $m^2 = 2H^2$, which is known to be 
a special case in 4D PF massive gravity \cite{Higuchi,DN}.
In the DGP model, there is a brane fluctuation mode with the 
same mass, which can have a mixing with the 
spin-2 graviton. We argued that this mixing presumably gives a ghost in the self-accelerating 
universe. Recently, a new instability is found in the limit of  
vanishing cosmological constant in the analysis of the 
gravitational shock wave fields generated by a source on the brane 
\cite{Kaloper}. This may be a signal of the appearance of the physical 
ghost in this limit.
In order to verify the existence of the ghost in this special case
rigorously, a more careful calculation of the effective action is 
required and this will be presented in a future publication \cite{future}.

We would like to thank C. Deffayet, N. Kaloper, K. Koyama, 
D. Langlois, R. Maartens and M. Sasaki for discussions. 
We would also like to thank D.S. Gorbunov and S.M. Sibiryakov 
for pointing out the existence of a solution for the spin-0
mode in the self-accelerating universe
and R. Rattazzi for pointing out the possibility of having a ghost 
from a mixing between the brane fluctuation mode and the spin-2 
graviton in a self-accelerating universe.
The author benefitted from discussions during the conference "The Next Chapter in 
Einstein's Legacy" at the Yukawa institute, Kyoto and 
the subsequent workshop. 
This work was supported by JSPS and PPARC.


\begin{references}
\bibitem{comment}
For $2/3< Hr_c < 1$, the mass is in the range $2H^2 < m^2 < 9H^2/4$
and for $Hr_c \leq 2/3$, there is no normalisable discrete mode.
\end{references}
\end{document}